%% file: main.tex
\documentclass[10.75pt, a4paper]{article}
\input{initialisation.tex}
\usepackage[framemethod=tikz]{mdframed}
\usepackage{lipsum}
\usepackage{enumitem}
\usepackage{mdframed}
\usepackage{physics}

\usepackage{setspace}
\usepackage{float}
\usepackage{subcaption}

\usepackage{subfigure}
\usepackage{graphicx} % Required for inserting images
\usepackage{amsmath}
\usepackage{authblk}
\usepackage[font={footnotesize,bf}]{caption}
\usepackage{caption,subcaption,graphicx}

\usepackage{float}

\usepackage[most]{tcolorbox}
\newtcolorbox{myBox}[3][]{
arc=2mm,
lower separated=true,
fonttitle=\bfseries,
colbacktitle=gray!10,
coltitle=black!50!black,
enhanced,
colframe=gray!10,
colback=gray!10,
title=#2,#1}

\usepackage{dirtytalk}
\usepackage{tcolorbox}
\usepackage[font=footnotesize]{caption}
\newtcolorbox{mybox}[1]{colback=green!6!white,colframe=black!75!black,fonttitle=\bfseries,title=#1}
\newtcolorbox{mybox2}{colback=red!5!white,colframe=red!75!black}

\usepackage{pifont}

\renewcommand{\vec}[1]{{\bm #1}}

\renewcommand{\grad}{\bm{\nabla}}
\newcommand{\diver}{\bm{\nabla}\!\cdot}
\newcommand{\rot}{\bm{\nabla}\!\times}

\newcommand{\Oe}{\mathrm{Oh}^{\mathcal{E}}}
\newcommand{\Oo}{\mathrm{Oh}^{\mathcal{O}}}

\newcommand{\even}{\mathcal{E}}
\newcommand{\odd}{\mathcal{O}}

\usepackage{soul,xcolor}
\setstcolor{red}

\newcommand{\pderiv}[2]{\dfrac{\partial#1}{\partial#2}}
\newcommand{\ptderiv}[1]{\pderiv{#1}{t}}

\usepackage{xcolor,hyperref}
\hypersetup{
   colorlinks,
   linkcolor={blue!50!black},%{red!80!black},
   citecolor={blue!50!black},
   urlcolor={blue!80!black}
} 
\definecolor{mycolor}{rgb}{0.122, 0.435, 0.698}

\usepackage{textgreek}
\usepackage{multicol}
\title{Non-Reciprocal Capillary Waves}
\newcommand{\equalcontrib}{\textsuperscript{\S}}
\usepackage{esint}

\author[1]{Holly du Plessis\equalcontrib}
\author[1]{Pedro Cosme\equalcontrib\footnote{p.a.cosmeesilva@uva.nl, ORCID: 0000-0001-6754-7762}} 
\author[1]{Hugo França\footnote{ORCID: 0000-0002-5361-7704}}
\author[1,2]{Maziyar Jalaal\footnote{m.jalaal@uva.nl, m.jalaal@damtp.cam.ac.uk, ORCID: 0000-0002-5654-8505}}
\affil[1]{Van der Waals-Zeeman Institute, Institute of Physics, University of Amsterdam, Amsterdam, The Netherlands}
\date{June 2025}
\affil[2]{Department of Applied Mathematics and Theoretical Physics, \protect\\
University of Cambridge, Wilberforce Road, Cambridge CB3 0WA, United Kingdom}

\begin{document}

\input{colors_define.tex}
\begingroup
\sffamily
\date{}
\maketitle

\begingroup
\renewcommand\thefootnote{\S}
\footnotetext{contributed equally}
\endgroup

\endgroup

\begin{abstract}
Capillary waves are a classical free-surface phenomenon in fluid mechanics, yet their behavior in chiral fluids remains largely unexplored. We show that odd viscosity breaks the reciprocity of capillary waves. Using linear theory together with fully nonlinear direct numerical simulations, we find that surface tension creates two inequivalent branches of odd capillary waves: a dispersive branch and a quasi-acoustic branch absent in the capillarity-free limit. Their unequal propagation and attenuation transform standing waves into traveling waves and produce an anomalously deep vortical boundary layer. Above a threshold odd viscosity, nonlinear accumulation of vorticity near the surface reverses the induced shear current and drives bulk particles opposite to the wave motion, giving rise to an anti-Stokes drift with no counterpart in conventional fluids. Our results show how combining capillarity with broken parity can be used to control wave propagation and transport at fluid interfaces, opening a route toward one-way fluidic waveguiding and chirality-programmed interfacial flows.

\end{abstract}

% \begin{multicols}{2}
\section*{Introduction}
\label{sec:introduction}
Capillary waves are the natural oscillations of a fluid interface when surface tension is the dominant restoring force. Beyond their classical role in short-wavelength free-surface dynamics, they provide a paradigmatic setting for strongly nonlinear interfacial phenomena, including solitary states, singular pinch-off events, and capillary-wave turbulence~\cite{Lioubashevski1996,Day1998,Falcon2007,Deike2014, jalaal2019capillary}. Controlling such waves is important for steering transport and deformation at interfaces and for enabling functionalities such as particle manipulation, patterning, wave refocusing, and and manufacturing with interfacial waves. Existing strategies typically rely on tailoring the effective material response (rheology) of the medium or interface~\cite{Wagner1999,Jiang2007,jalaal2019laser,jalaal2021spreading,sanjay2021bursting,Kharbedia2021,Shen2018}, or on shaping waves through spatial and spatiotemporal modulation~\cite{Francois2017,Bacot2016,Vidler2024}. Yet how to achieve robust \emph{intrinsic} directionality at a free surface remains an open question.
One promising route is to use fluids that break parity and time-reversal symmetry. In recent years, such symmetry-broken materials have moved from abstract theory to experimental reality, from chiral colloidal free-surface flows to living chiral crystals~\cite{soni2019odd,tan_odd_2022}, while theory has shown that odd mechanics is closely tied to nonreciprocity and can qualitatively reshape free-surface dynamics~\cite{AbanovMonteiro2019,MarkovichLubensky2024,Cosme2023a,PhysRevLett.122.128001,PhysRevLett.122.214505,veenstra2024non}.
Here, we show that odd viscosity provides a control knob for capillary waves, rendering the free surface chiral and nonreciprocal through unequal counter-propagating modes, anomalously deep vortical boundary layers, and reversed drift.

Odd viscosity is the non-dissipative, parity-breaking component of a fluid’s stress response, arising whenever microscopic dynamics violate time-reversal or mirror symmetry (\emph{e.g.}, via cyclotron motion in magnetized plasmas~\cite{PhysRevLett.75.697,Cosme2023a} or steady particle rotation in active colloids~\cite{soni2019odd,Yan2015}). In a two-dimensional incompressible fluid, its effect appears as an antisymmetric block in the rank-four viscosity tensor, orthogonal to the usual shear and bulk viscosities~\cite{Fruchart_2023,avron1998oddviscosity}. 
% -- and endows the flow with transverse stress proportional to local vorticity. 
In such a configuration, the contribution of odd viscosity is most pronounced when stress boundary conditions such as free surfaces exists, where uni-directional waves~\cite{markovich2024chiralactivefluidslearn}, stabilized thin films~\cite{stabalisation_2019}, transverse edge states~\cite{PhysRevLett.122.128001}, chiral bubble dynamics~\cite{Caprini2025,Guo2025}, lift~\cite{hosaka_hydrodynamic_2021}, symmetry-broken thermo-capillary action~\cite{thermo_odd_2022}, and modulated droplet motion~\cite{francca2025odd} have been observed. By engineering microscopic chirality one thus gains direct control over dispersion and nonlinear interactions at the boundary, making odd-viscous fluids a fertile platform for tunable, non-reciprocal wave phenomena.

While recent studies have examined odd‐viscosity surface waves either neglecting surface tension~\cite{PhysRevFluids.6.L092401,Abanov_2018} or focusing on linear, small‐amplitude perturbations~\cite{Granero_Belinch_n_2022}, the present work departs from these approaches by solving the full nonlinear Navier–Stokes equations via Direct Numerical Simulation (DNS) coupled with a Volume of Fluid (VOF) method to capture large interface deformations, thereby transcending the small‐amplitude regime. Such fully resolved flow simulations are essential for engineering chiral fluids with tailored, non‐reciprocal wave behavior. By resolving the fluid domain both above and below the interface, we uncover striking flow structures, most notably extended vorticity and an anomalous Stokes drift.

\begin{figure*}[h!]
    \centering
    \includegraphics[width=1\linewidth]{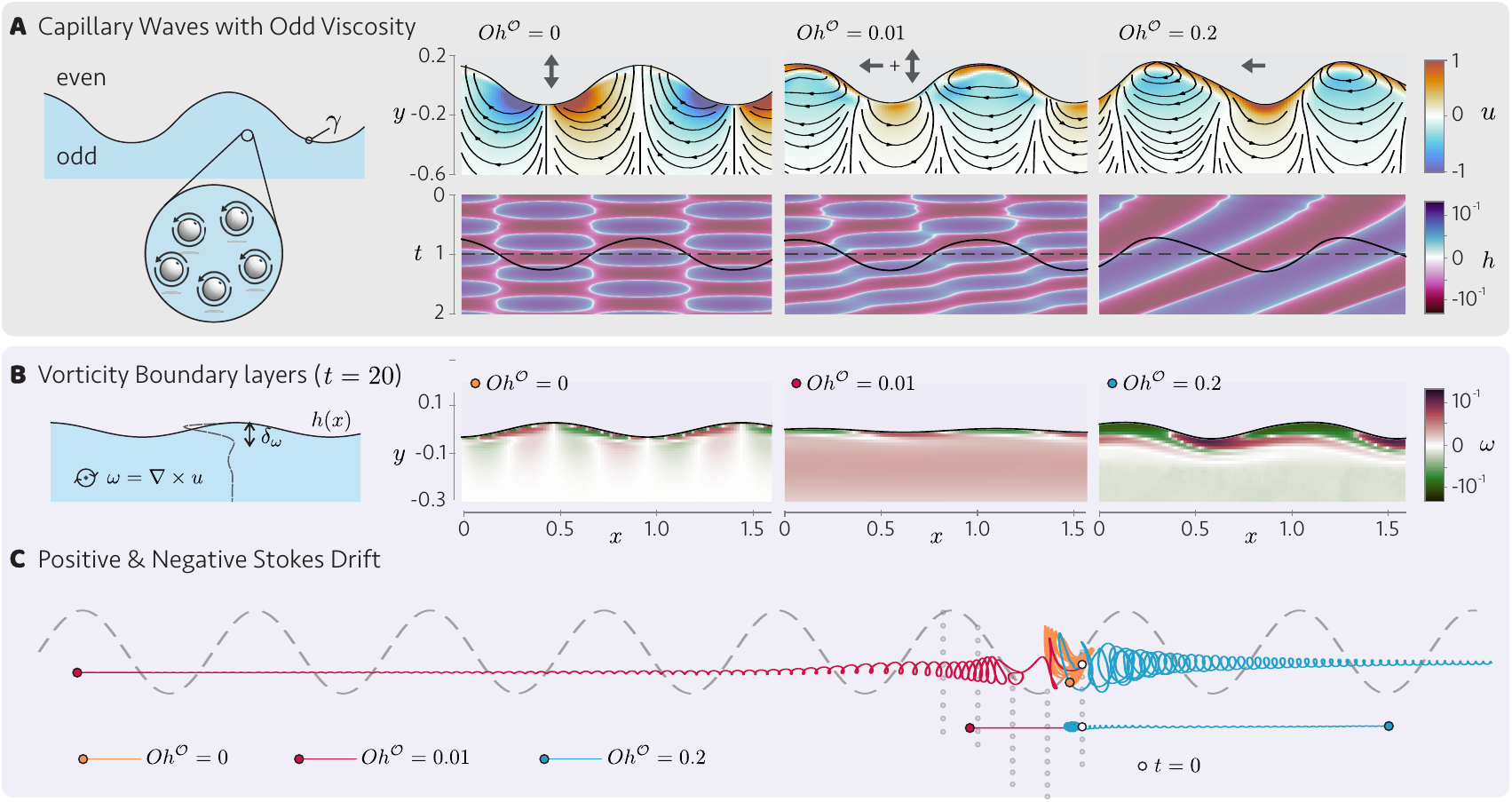}
%    \caption{Fig.1}

\caption{\textbf{Non-reciprocal capillary waves with odd viscosity.}
Dynamics of capillary waves with even shear viscosity $\Oe=0.001$ and amplitude $a=0.2\lambda$, shown for three simulations with $\Oo=0$, $\Oo=0.01$, and $\Oo=0.2$.
\textbf{(A)} Horizontal velocity field ($u$) at $t=1$ and spatiotemporal evolution of the interfacial height $h$ over $t\in[0,2]$. The onset of unidirectional wave motion is visible in the flow field, which exhibits a net velocity to the left.
\textbf{(B)} Vorticity field at $t=20$, highlighting the vortical boundary layers that form near the interface. The penetration depth of this layer increases markedly with odd viscosity, together with the associated shear current, leading to the emergence of Stokes-like and anti-Stokes-like drift.
\textbf{(C)} Lagrangian tracer particles placed inside the liquid. For each simulation, the trajectories of two particles are shown: one initially near the surface and one deeper in the fluid, below the boundary layer (initial positions marked by white circles). Depending on depth and odd viscosity, tracers drift either with the wave or against it.}

% which illustrates the horizontal velocity profile and streamlines at $t=1$ which illustrate uni-directional wave motion and the spacial temporal evolutions between $t=0$ to $t=2$ which show the development of traveling wave motion. In the later moments of (B) the vorticity profile at $t=20$ is illustrated and the trajectories of two tracer particles where one is initialized just below the surface and a second is entrenched deeper in the fluid below the vorticity boundary layer, this shows a change the boundary layers penetration depth with odd viscosity and a shear current which forms in both the favorable and adverse direction enabling particles to exhibit a stokes-like drift and anti-stokes drift against the wave motion.
%}
\label{fig:1}
\end{figure*}

Consider an incompressible 2D odd fluid flow subject to surface perturbations at a fluid-fluid interface (cf.~Fig.\ref{fig:1}A). The governing equations are an extension of Navier-Stokes equations, including odd contributions as well as capillary effects:
\begin{gather} 
\nabla \cdot \bm{u} = 0 \\
\mathrm{D}\,\bm{u}\, /\, \mathrm{D}\, t 	= \diver \vec{\sigma} + \bm{F}^{\mathcal{S}}
\label{eq:nondim_navier}
\end{gather}
Here, $\mathrm{D}\,\bm{u}\, /\, \mathrm{D}\, t$  is the material derivative, and $\vec{u} (\vec{x},t)$ is the velocity vector. The stress tensor $\vec{\sigma} = p\vec{I} +\vec{\sigma}^e + \vec{\sigma}^o$ comprises conventional (even) term $\grad \cdot\vec{\sigma}^e = \Oe\,\grad^2 \vec{u}$ and the odd one $\grad \cdot\vec{\sigma}^e = \Oo\, \vec{\epsilon} \,\grad^2 \vec{u}$, where $\Oe$ and $\Oo$ are even and odd Ohnesorge numbers, here can been seen as normalized shear and odd viscosities, respectively (see appendix~\ref{app:EqsandNumerics} for the details of equations and non-dimensionalisation). Note that, $\vec{\sigma}^o$ uses the Levi-Civita symbol $\vec{\epsilon}$ to rotate the velocity‐gradient before symmetrizing and can also be written as $\diver \vec{\sigma}^o = \Oo \grad \omega $, where $\omega=(\rot\vec{u})\cdot\hat{\vec{z}}$ is the vorticity. Importantly, given that the odd contribution can be written as the gradient of a pseudoscalar, one can absorb it into a redefinition of the pressure $p\mapsto p+ \Oo\,\omega$, rendering the fluid equations identical to the fully even case~\cite{avron1998oddviscosity,Fruchart_2023}.
$\bm{F}^{\mathcal{S}} = \kappa \, \delta_s \, \bm{n}$, accounts for the normalized surface tension forces, linearly correlated with the interfacial curvature $\kappa$ and centered on the interface with the Dirac function $\delta_s$ (again as detailed in appendix~\ref{app:EqsandNumerics}).
Gravity is ignored in the analysis and simulations presented here (see appendix~\ref{app:Dispersion}).

% We use a combination of Finite Volume and Volume of Fluid (VoF) Methods~\cite{Hirt1981,popinet2009accurate,Popinet2015}, implemented in the open-source code Basilisk~\cite{Popinet2013-Basilisk} to solve these equations (see appendix B ). 

% \kappa \, \delta_s \, \bm{n} \right) / We$. Here, $\kappa$ is the curvature of the interface between $\Omega_1$ and $\Omega_2$, and $\bm{n}$ is the normal vector to that interface. 

% The superscripts $e$ and $o$, denote the 
% The second term on the right-hand side is the divergence of the  deviatoric stress tensor $\bm{\nabla} \cdot \bm{\tau}^e = \left(  \bm \right)$
% The dynamics of the surface is governed by the balance of surface tension and internal stress; 
% but, for odd stresses to manifest, it is required that a clear axial direction exists. In the present work, we take that the surface height perturbations (regarded as along $y$) only propagate in one direction (taken as $x$), such that, by symmetry, the problem reduces to an effective two-dimensional (2D) scenario with the direction of the vorticity and internal angular momentum lies perpendicular to both wave vector and surface normal (cf.~Fig.\ref{fig:1}). 

\section*{ 
Chiral Non‑Reciprocal Capillary Waves}

\begin{figure*}[t]
    \centering
    \includegraphics[width=1\linewidth]{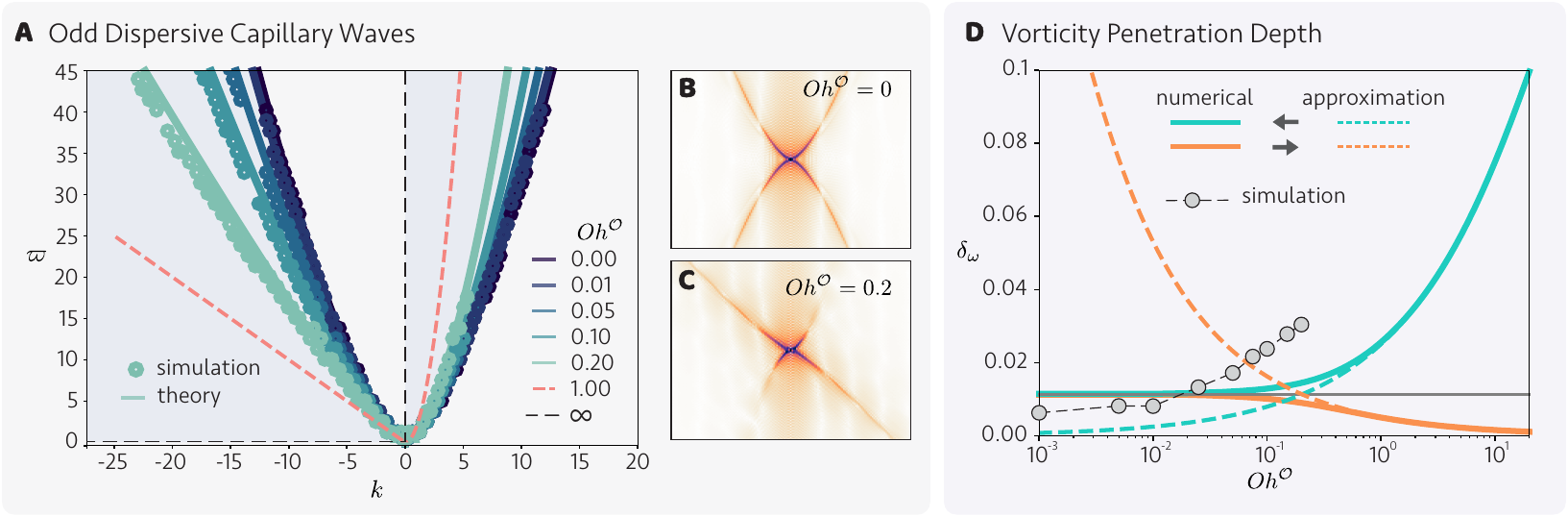}
\caption{\textbf{Dispersion branches and penetration depth.}
\textbf{(A)} Dispersion branches for odd capillary waves, extracted from simulations initialized with a boxcar-like perturbation of width $w=1$ and height $h=0.1$ by identifying the peaks of the Fourier-transformed signal.
\textbf{(B,C)} Examples of the corresponding spectra in Fourier space for $\Oo=0$ and $\Oo=0.2$, respectively. With increasing odd viscosity, the dispersion branches become progressively asymmetric, approaching a linear negative branch and a quadratic positive branch when $\Oo\sim 1$.
\textbf{(D)} Penetration depth measured from low-amplitude sinusoidal waves with $a=0.05$, using the characteristic decay length of the vorticity, defined as the depth at which the vorticity falls to $1/e$ of its surface value. The measurements are compared with the linear prediction $m^{-1}$; when the negative mode dominates, the data follow the trend of the diverging penetration depth. The measured decay length is nevertheless larger than predicted by linear theory.}   
\label{fig:2}
\end{figure*}

\textbf{Linear Theory.}
The fortunate circumstance that the odd viscous contribution can be cast as a vorticity-corrected pressure allows us to resort to the well-known Lamb solutions~\cite{lamb1975hydrodynamics,} when studying the linear regime of our perturbed system (see appendix~\ref{app:Dispersion} for details), leading to the dispersion relation:
% 
% In combination with the linearized  DBC, which reads
% \begin{subequations}
%     \begin{align}
%     \eta^e(\partial_y u+\partial_x v)+2\eta^o\partial_x u&=0\,\,\text{and}\\
%     -(p-\eta^o\omega)+2\eta^e\partial_y v+2\eta^o\partial_x v&=\gamma\rho \partial_{xx}h,
% \end{align}
% \end{subequations}
% such linear solutions of the fluid domain lead to the dispersion relation 
\begin{equation}
    \begin{vmatrix}
 \varpi^2+2 \, ik \, \big( \Oe  \, k-i  \, \Oo|k|\big) \, \varpi -|k|^3  & 2 \, ik \, \big( \Oe  \,  m \, \operatorname{sgn}k-i  \, \Oo|k|\big) \, \varpi  -|k|^3\\[1em]
  2 \, ik \, \big( \Oe \,  k-i  \, \Oo |k|\big)  & \varpi   + 2 \, ik \, \big( \Oe k-i \,  \Oo m\big)
    \end{vmatrix}=0,\label{eq:full_dispersion}
\end{equation}
where 
% we have discarded the influence of gravity, and 
% made use of the shorthand 
$\varpi$ and $k$ are frequency and wavenumber, respectively, and where we used the shorthand $m^2=k^2-i\varpi/\Oe$. 
% In the previous equation and hereafter, we define the even and odd Ohnesorge numbers respectively as  $\Oe=\eta^e/\sqrt{\rho^2\gamma L}$ and $\Oo=\eta^o/\sqrt{\rho^2\gamma L}$ while scaling the frequencies $\varpi$ and time with the characteristic time scale $\sqrt{L^3/\gamma}$ dictated by the surface tension and where the length scale $L$ is taken to be a relevant length, mainly the wavelength of the initial condition. 
% 
Note that in the fully inviscid limit of $\Oe\to0$ and $\Oo\to0$, we recover the known capillary dispersion $\varpi^2=|k|^3$~\cite{lamb1975hydrodynamics,landau1987fluid,lannes2013water}. However, if instead, we take the limit of zero shear viscosity while retaining its odd counterpart (that is to say $\Oo\gg\Oe$), we have
\begin{equation}
  \varpi^2 +2 \, \Oo \, k|k| \, \varpi -|k|^3  =0,
\end{equation}
with a cross term that breaks the symmetry of the roots.
Thus elucidating the role of the odd viscosity as an explicitly chiral or non-reciprocal term that tilts and distorts the classical dispersion of surface waves. Specifically, it renders the left and right propagating modes unequivalent and  $\varpi(-k)\neq\varpi(k)$ for any branch of the dispersion, such that in a waveguide or periodic domain, one of the chiralities overtakes the other. This is clearly observed on our simulations (\emph{cf.} Fig.~\ref{fig:1}A) where, when odd viscosity is introduced, a travelling wave develops. Such a wave is born out of the competition between the unequivalent left and right-propagating modes.
Importantly, note that this asymmetry in $k$-space is beyond any simple boost since it alters the group velocity in a nonconstant way, as seen in Fig.\ref{fig:2}, it tends to bring one of the branches towards an acoustic limit with $\varpi\simeq k/2 \Oo$ while the second one approaches a quadratic dispersion $\varpi\simeq-2 \, \Oo \, k|k|$. In fact, this latter branch is the only one present in the absence of surface tension, addressed in previous works \cite{PhysRevFluids.6.L092401,Abanov_2018}, whereas our approach indicates that an additional mode is allowed for fluids exhibiting surface tension; this is thus particularly important for low odd viscosity scenarios, given the $1/2\,\Oo$ divergence. 

From our simulations, we report that, while an initial sinusoidal perturbation evolves, as expected, to standing waves if $\Oo=0$, by the inclusion of odd viscosity, the parity breaking gives rise to a net displacement. This transition between a fluid with no odd viscosity $\Oo=0$ to fluids with non-zero odd viscosity is illustrated in Fig.~\ref{fig:1} where the kymograph of an even fluid shows the spacial-temporal evolution consistent with a standing wave, whereas the odd waves develop a traveling wave which captures the dominance of the chiral mode. The development of the uni-directional wave emerges faster in systems with a larger odd shear viscosity as the disparity in the attenuation between left and right movers also increases. One can understand such behaviour as the result of the normal stress, exerted on the interface, that is converted into kinetic energy, being now coupled to the shear modes due to odd viscosity, thus generating transverse motion instead of oscillatory dynamics, as transverse flows cannot be converted back into the energy of the interface. 

We retrieved the dispersion of odd waves from numerical simulations by taking the space-time Fourier transform of a surface perturbation with a rich spectral content, such as a boxcar function. Such Fourier transforms are given in Fig.~\ref{fig:2} for $\Oo=0$ and $\Oo=0.2$, which reflects the symmetric dispersion of an even fluid in (B) and the development of a dominant quasi-acoustic mode in (C). The formation of this quasi-acoustic mode can be seen in panel (A) of Fig.~\ref{fig:2} as an increase in the shear odd viscosity produces a larger distortion of the dispersion towards the left, which is well aligned with the theory. As the odd viscosity increases, the dispersion relation deviates more quickly from linear theory, which is consistent with observations where cases with larger odd viscosity demonstrate enhanced nonlinear properties triggered by the stronger dispersive nature.

\pagebreak
\textbf{Vortical Boundary Layer.} Taking the limit of $\Oe\to 0$ restrains us to situations where there is no attenuation of the wave solutions. However, the imaginary part is central not only for the decay but also for the penetration depth of the vortical boundary layer $\delta_\omega\sim \Re (m)^{-1}$. In fact, the vorticity of the linear solutions follows $\omega \propto e^{i k x-i \varpi t}e^{m y} \, (i \varpi / \Oe)$,
% \begin{equation}
% \omega \propto e^{i k x-i \varpi t}e^{m y}\frac{i \varpi}{\Oe} ,
% \end{equation}
decaying away from the interface 
and here the different nature of the two modes is key. It can be shown that
for $\Oe\ll \Oo\sim 1 $ the dispersion branches are well approximated by   
\begin{equation}
    \varpi=\frac{k}{2\,\Oo}-i \frac{\Oe}{{\Oo}^2}\,k \quad\text{and}\quad \varpi=-2\,\Oo \, k|k|-i\sqrt{\Oo \Oe} \,k^2. \label{eq:approximation_dispersion}
\end{equation}
As such, while for the dispersive mode the penetration length is monotonically decreasing and $ \delta_\omega\xrightarrow{\Oo\to\infty}0 $
% \begin{equation}
%     \delta\xrightarrow{\Oo\to\infty}0 
% \end{equation}
in the case of the quasi-acoustic mode it is monotonically increasing and $\delta_\omega\xrightarrow{\Oo\to\infty}1 $.
% \begin{equation}
% \delta\xrightarrow{\Oo\to\infty}1 
% \end{equation}

This phenomenon, only possible when surface tension is included, can be clearly observed in our simulations, as shown in Fig.~\ref{fig:1}B for increasing odd Ohnesorge numbers. Also, it will be central to the ensuing discussion of near-interface drifts (\emph{cf.} Fig.~\ref{fig:1}C). Yet, while the linear theory provides us with the correct behaviour and the retrieved numerical dispersion relation agrees closely with the theory for the real part of $\varpi$, as can be seen in Fig.~\ref{fig:2}A. The scaling of $\delta_\omega$ from both the approximations \eqref{eq:approximation_dispersion} or even \eqref{eq:full_dispersion} seems to underestimate the simulated penetration depth, as illustrated in Fig.~\ref{fig:2}D, even though yielding the correct overall trend. We attribute such an offset to the nonlinear contributions that the preceding theory does not cover. Thus, the DNS approach to the problem is critical for going beyond the linear theory and for accurately describing the phenomena triggered by the interplay of odd viscosity and surface tension, where nonlinearity is manifestly important. 

%\textcolor{red}{say what all these means - talk about fig 1 -  connect to DNS say simulations theory agree  buuut nonlinearity is important -> we gonna test these and go beyond}

% \section{Results and discussion} 
% \subsection{Features of odd capillary waves}

Although the mode difference cannot be simply seen as a boost, the unidirectional wave motion is qualitatively similar and can thus be compared to progressive waves seen in classical fluids driven by a surface wind. %where streamlines which form around the crest and troughs of the wave. 
In that scenario, the flow structure is consistent with mass transport classically described by Stokes drift theory \cite{Stokes1847,vandenBremer2017,vandenBremer2019}, the situation in the odd fluids is, however, quite different. In the odd waves, closed streamlines form beneath the crests of the waves (cf. Fig. ~\ref{fig:1}), unlike even fluids due to the emergence of a boundary flow. 
%
%This flow emerges from the odd viscosity modification of the tangential boundary condition to induce an effective no-slip condition when the ratio of the odd to even shear viscosities is large ~\cite{}. The emergence of this tangential flow is more prevalent at higher amplitudes as odd non-linear terms have a stronger signature, as expected for effects tied to a boundary layer, inherently nonlinear in nature. 
%
On longer timescales, once the system has time to equilibrate, the edge flow vanishes, but an increased vorticity boundary layer remains as a signature of the odd viscosity. 
Additionally, when inertia is not negligible, vorticity is convected away from the surface, leading to the formation of a shear current that induces particle drift velocities much larger than those prescribed by Stokes drift in irrotational and non-inertial waves. %This shear layer conventionally has the opposite sign to the direction of wave propagation ~\cite{}. 

\begin{figure}[!ht]
    \centering
    \includegraphics[width=0.76\linewidth]{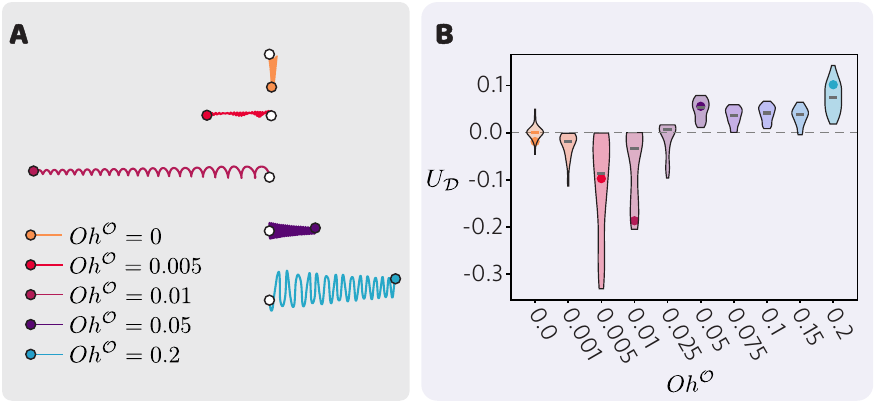}
\caption{\textbf{Tracer drift and drift inversion.}
\textbf{(A)} Drift of a tracer particle in the bulk for increasing odd viscosity, showing the inversion from leftward to rightward motion at large $\Oo$. All particles are initially positioned at the same initial relative height from the interface.
\textbf{(B)} Drift-velocity distribution and the average velocity obtained from 50 tracer particles placed across a full wave-length and at different relative depths of $0$ to $-0.5$ from the interface and temprally averaged at late time of $t \in[8,10]$, showing the transition from conventional drift ($U_D<0$) to anomalous drift ($U_D>0$) at high $\Oo$. The colored circles highlight the examples shown in panel A.}   
\label{fig:3}
\end{figure}

\textbf{Anomalous Drift \& Highly Non-linear Waves.} While in conventional fluids the drift is dictated, and follows, the \emph{driving wind}, odd fluids with surface tension present two distinct regimes. % 
In the case where the odd viscosity is larger than shear viscosity but still much smaller than the effect of surface tension ($\Oo\ll 1$), a shear current can be seen forming that is aligned with the dynamics characteristic of classical waves. In this regime, even though parity is broken, leading to unidirectional wave motion, the effects of odd viscosity are still small and can still be approximated by classical dynamics by an equivalent travelling wave on an even fluid. That is to say, the drift of particles in the bulk follows the travelling wave direction; in our frame of reference this is towards the left and the drift velocity $U_D<0$. A behaviour that we observe in the simulations, as shown in Fig.~\ref{fig:3}. 

\begin{figure*}[!ht]
    \centering
    \includegraphics[width=1\linewidth]{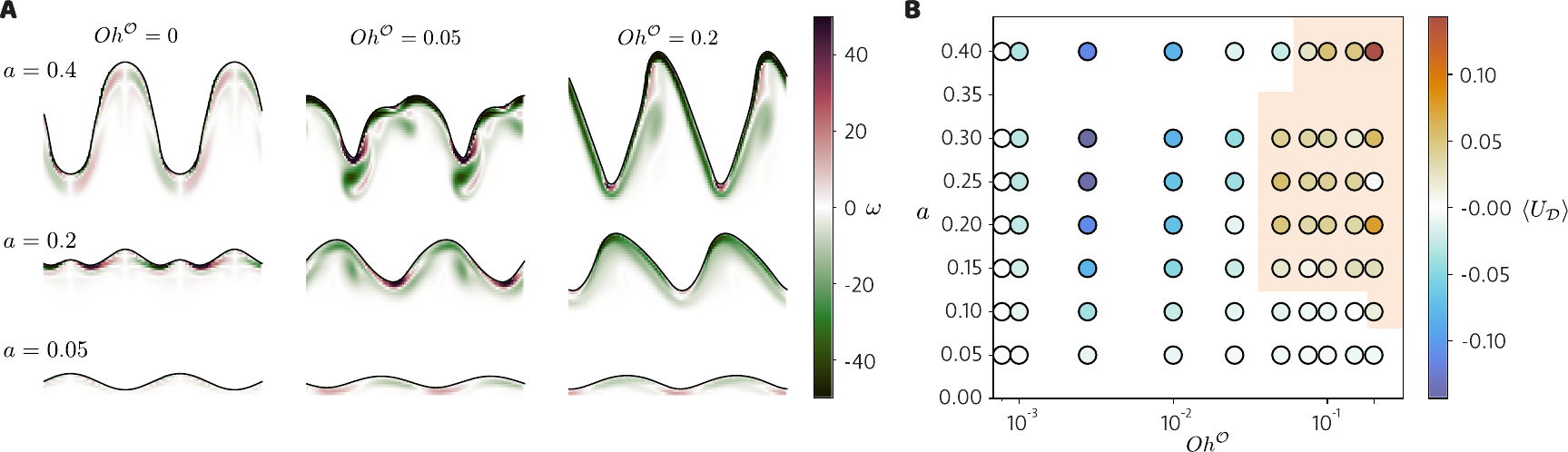}
\caption{\textbf{Nonlinear non-reciprocal waves.}
\textbf{(A)} Influence of wave initial amplitude, $a$, and odd viscosity, $\Oo$, on the waveform and vortical boundary layer. Increasing nonlinearity together with odd viscosity steepens the wave and enhances the concentration of vorticity near the surface.
\textbf{(B)} Parameter map of the mean drift velocity as a function of wave amplitude, $a$, and odd viscosity, $\Oo$. The drift inversion is absent in the weakly nonlinear regime ($a\ll 1$) and, once it appears, its threshold is governed primarily by $\Oo$.}  
\label{fig:4}
\end{figure*}

However, in the opposite regime,  where odd viscosity dominates, and similarly, it is closer to surface tension ($\Oo\lesssim 1$), the fluid develops a stronger quasi-acoustic mode, as described previously when discussing Eq.~\eqref {eq:approximation_dispersion}. We then observe dynamics atypical of travelling waves with the transition to an anti-Stokes drift, where bulk tracer particles are transported in the opposite direction of the travelling surface wave. This transition seems to occur for $\Oo\approx0.03$ (\emph{cf.} Fig.~\ref{fig:3}B and Fig.~\ref{fig:4}B). Curiously, this coincides with the value of $\Oo$ where we observe the penetration depth from the simulations to cross the limiting theory of $\Oo\sim0$ (\emph{cf.} Fig.~\ref{fig:2}D). Therefore, we conjecture that the mechanism of anti-Stokes drift is triggered by the nonlinearities of the system. In fact, with higher amplitudes, the interface develops wave profiles with large curvature and produces a shear current that changes sign due to large regions of negative vorticity forming at the surface, which is concentrated in the troughs of the wave. 

Such observation and proposed connection to nonlinearity prompted us to probe the effects of wave amplitude on the dynamics of the system. Indeed, increasing the amplitude of the initial condition in combination with increasing odd viscosity has several repercussions both on the flow and the free surface. Namely, with the steepening and asymmetry of the waveform, but critically with a sharper, more localised, vorticity penetration layer Fig.~\ref{fig:4}A. In turn, this accumulation of vorticity then leads to the observed contrary drift. The hypothesis that we are in the face of an inherently nonlinear phenomenon is corroborated by the fact that it vanishes for small enough amplitudes, as shown in Fig.~\ref{fig:4}B. Yet, the gathered data indicate that, once present, the transition to inverse drift is mostly determined by the value of odd viscosity, with the threshold aforementioned. Further DNS with a wider and more refined parameter sweep in the future shall certainly clarify the details of the transition, but our results already show a clear trend. 

Still, to further elucidate the interplay between nonlinearities and the dispersive nature of odd viscous fluids, a reduced-dimensional nonlinear model is necessary; similar to what was obtained in similar odd viscous systems ~\cite{Cosme2023a}. While the development of such a theory would go beyond the scope of the present work, the weakly-nonlinear regime might already shed some light. Indeed, from the structure of the dispersion relation, one can construct a scalar theory for nonlinear capillary odd waves as (see appendix~\ref{app:NonlinearModel})
\begin{equation}
    \phi_{tt}+(s \phi \phi_x)_t-\operatorname{H}\left[2\Oo \phi_{xxt}+\phi_{xxx}\right]=0\label{eq:nonlinear2ndBO}
\end{equation}
where $\operatorname{H}[\cdot]$ is the Hilbert transform, defined as the multiplier operator with symbol $-i\operatorname{sgn}(k)$  ~\cite{HilbertTransforms2009,Dahne2023} and $\lambda$ a coefficient to be determined. This generalises Benjamin--Ono equation (BO) to second order. Like BO, it admits algebraic solitons as travelling solutions for constant $\lambda$, of the form
\begin{equation}   
\phi(x-ct)\equiv\phi(\xi)=\frac{4c(1-2\Oo c)^2{s^{-1}}}{c^4\xi^2+(1-2\Oo c)^2}
\end{equation}
already showcasing that there exist critical combinations of propagation velocity $c$ and odd Ohnesorge that control the nature of the perturbation. 

\textbf{Conclusion.} Capillary waves in an odd fluid define a qualitatively new interfacial dynamics. Surface tension splits the response into two inequivalent branches (a dispersive mode and a quasi-acoustic mode) whose unbalanced competition breaks reciprocity and converts nominally standing perturbations into travelling waves. Odd viscosity also qualitatively reshapes the near-surface flow: while the dispersive branch remains shallow, the quasi-acoustic branch exhibits a vortical boundary layer whose penetration depth grows with the odd Ohnesorge number, in contrast to the usual intuition from viscous boundary layers. Above a critical odd viscosity, nonlinear accumulation and advection of vorticity near the interface reverse the induced shear current and generate an anti-Stokes drift, so that bulk tracers move against the direction of wave propagation. These findings identify odd capillary waves as a minimal free-surface setting in which broken parity converts interfacial oscillations into directional transport, opening a route to chirality-programmed transfer of tracers, momentum, and potentially wave energy along fluid interfaces. In this sense, the interface can act as a fluidic conveyor belt, echoing broader non-equilibrium mechanisms of directed transport in pattern-forming systems. An immediate challenge is to derive a nonlinear theory for the drift inversion and mode competition, and to extend the framework to fully chiral fluids in which intrinsic angular-momentum transport and free-surface stresses are treated on equal footing.

% \begin{itemize}
%     \item odd capillary waves have two unequivalent modes
%     \item making it nonreciprocal due to unbalanced competition of the modes
%     \item New behavour of the penetration depht of the vorticity :  growing with $\Oo$  
%     \item above a critiral $\Oo$ such layer induces a inverse drift 
%     \item nonlinear behaviour is key to these phenomena
%         \item directed mass/energy  transport ? 
%     \item tie it all with Erwin Freys conveyor belts?
% \item future work:  nonlinear theory
% \item future work: fully chiral fluid ang. momentum transport
% \end{itemize}

\section*{Acknowledgements}
The authors thank Corentin Coulais, and Sander Mann for discussions. M.J. acknowledges the ERC grant no.~``2023-StG-101117025, FluMAB" and the Vidi project Living Levers with No. 279 21239, financed by the Dutch Research Council (NWO).

\printbibliography
%\bibliography{bibliography}

% \clearpage
% \renewcommand\thefigure{S\arabic{figure}}
% \setcounter{figure}{0}    
% \section*{Supplementary Information}
% \subsection*{Supplementary Movies}
% Supplementary ...\\

\pagebreak
\appendix
\renewcommand{\thesection}{\Alph{section}}
\titleformat{\section}
  {\bf\sffamily}
  {APPENDIX \thesection:}
  {5pt}
  {\MakeUppercase}

\section{Governing Equations \& Numerical Implementation}
\label{app:EqsandNumerics}
The stress tensor of twodimensional incompressible fluids with broken parity can be given as $\vec{\sigma} = p\vec{I} +\vec{\sigma}^e + \vec{\sigma}^o$. With its deviatoric part comprised of both the conventional term proportional to the strain rate and the odd one, that scales with the rotated velocity gradient as:   
\begin{align}
    \vec{\sigma}^e&=\eta^e     \begin{bmatrix}
      2\partial_x u & \partial_y u+\partial_x v \\
     \partial_y u+\partial_x v&2\partial_y v   
     \end{bmatrix} ,%\triangleq \eta^e \dot{\nu},
     \\
    \vec{\sigma}^o&=\eta^o     \begin{bmatrix}
     -\partial_y u-\partial_x v &  \partial_x u-\partial_y v\\
     \partial_x u-\partial_y v& \partial_y u+\partial_x v 
     \end{bmatrix}, %\triangleq \eta^o\tilde{\nu},
     \label{eq:odd_stress}
\end{align} 
where the velocity field is written as $\vec{u}=u\hat{\vec{x}}+v\hat{\vec{y}}$.
Note that although both components are symmetric at the stress tensor level, the odd part can be traced back to the antisymmetric part of the viscosity tensor.
% \begin{comment}
% {\color{red}
% with the symmetric and asymmetric stress components expressed independently as
% \begin{align}
%     \sigma^e&= \eta^e \dot{\nu},\\
%     \sigma^o &= \eta^o\tilde{\nu} 
% \end{align} 
% where $\eta_e,\eta_o$ are the even and odd shear viscosities, $\dot{\nu}$ the strain rate and $\tilde{\nu}$ the odd strain given by 
% \begin{align}
%     \tilde{\nu} =
%     \begin{bmatrix}
%      -(\frac{\partial u}{\partial y}+\frac{\partial v}{\partial x}) &  (\frac{\partial u}{\partial x}-\frac{\partial v}{\partial y})\\
%      (\frac{\partial u}{\partial x}-\frac{\partial v}{\partial y})& (\frac{\partial u}{\partial y}+\frac{\partial v}{\partial x}) 
%      \end{bmatrix}
% \end{align}
% }
% \end{comment}
% 
From Cauchy's stress equation $ D_t\vec{u}= \diver \vec{\sigma}$, the odd term propagates through to Navier--Stokes as an odd force, $F^{odd}$, which in incompressible fluids can be cast as
\begin{align}
\diver \vec{\sigma}^o = \eta^o \grad \omega \label{eqn: odd force}
\end{align}
where $\omega=(\rot\vec{u})\cdot\hat{\vec{z}}$ is the vorticity. 
% Note that, given that the odd contribution can be written as the gradient of a pseudoscalar one can absorb it into a redefinition of the pressure $p\mapsto p+\eta^o\omega$, rendering the fluid equations identical to the fully even case.
The expression above can be equivalently expressed as $\eta^o\vec{\epsilon} \nabla^2 \vec{u}$ where $\epsilon$ is the Levi--Civita symbol, expressing the coupling between the shear and normal stresses due to odd viscosity. 

% As odd viscosity can be captured by a corrected pressure term we assume odd viscosity does not explicitly modify the boundary conditions. We maintain the conventional free surface boundary of
% \begin{align}
%     n\sigma n = \gamma \kappa,\quad t\sigma n = 0, \label{eqn:stress bc} 
% \end{align}
% where $\gamma$ is the surface tension coefficient and $\kappa$ the curvature of the interface ~\ref{}.
% }

The ability to write the dynamics of an incompressible odd fluid as a corrected even one allows us to apply the conventional boundary conditions, \emph{viz.} the kinematic boundary condition (KBC)
\begin{subequations}
\begin{equation}
    D_th=0,
\end{equation}
for the surface perturbation profile height $h(x,t)$, and the dynamic boundary condition (DBC) 
\begin{equation}   \vec{\sigma}\cdot\hat{\vec{n}}=\gamma\rho\,\kappa\hat{\vec{n}},
\end{equation}
\end{subequations}
with $\kappa$ the curvature of the interfacial surface and $\hat{\vec{n}}$ its normal unit vector and where $\gamma\rho$ stands for the surface tension coefficient.

\subsection*{Numerical implementation and non-dimensionalisation}

\begin{figure}
    \centering
    \includegraphics[width=0.7\linewidth]{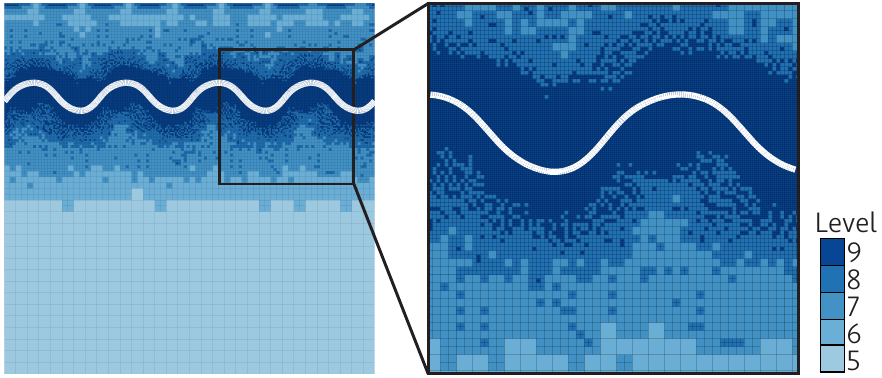}
    \caption{An example of adaptive mesh refinement near the interface.}
    \label{fig:mesh}
\end{figure}

The equations of motion are solved using the open source code Basilisk~\cite{PopinetBasiliskC}, using its Navier--Stokes and Volume-of-Fluid (VOF) solvers~\cite{popinet2009accurate,popinet2018numerical} together with our implementation of odd-viscous stress module (see below). The code uses adaptive cartesian grids (see Fig.~\ref{fig:mesh}) to discretize the partial differential equations, allowing increased numerical resolution where required, such as near the deformable interface. 
% A detailed description of the can be accessed at~\cite{} as in the following section only a brief overview is given.
% 

We use a single-fluid model in which the two phases are treated as a single effective fluid with spatially varying density and viscosity functions. The model uses a scalar color function $c(\vec{x},t)$ to indicate the fraction of a given fluid contained in each numerical cell. The density and viscosities are then locally calculated as $\rho (c) =  c \ \rho_w + (1 - c)\rho_a$, $\mu^\even (c) = c \  \mu_w^\even + (1 - c)\mu_a^\even$, and\ $\mu^\odd(c) = c \  \mu_w^\odd$. Throughout this paper, we take the properties of the ambient fluid as $\rho_a = 10^{-2}\rho_w$ and $\mu_a^\even = 10^{-2}\mu_w^\even$.

The odd viscosity term is implemented by modifying the standard Basilisk ``viscosity.h" module and adding the new terms corresponding to the odd part of the viscous stress tensor given in Eq.~\eqref{eq:odd_stress}. This approaches guarantees that the viscosity term is still discretized implicitly within the solver, which leads to better numerical stability.

% As odd viscosity breaks macroscopic symmetries the evaluation of Navier-Stokes equations in each dimension have to evaluated separately.

% \begin{align}
%    \frac{\partial u}{\partial t}+u \cdot \nabla u  = \frac{1}{\rho}(- \nabla p +\nabla \cdot \bar{\bar{\sigma^e}} + \nabla \cdot \bar{\bar{\sigma^o}} + f)
% \end{align}
% Where f is the body force and $\nabla \cdot  \bar{\bar{\sigma^o}}$ is
% \begin{align}
%     \nabla \cdot  \bar{\bar{\sigma^o}} = \begin{bmatrix}
% \eta_o ( \partial_x \partial_y v_x -\partial_x^2 v_y + \partial_y \partial_x v_x -  \partial_y^2 v_y) \\
% \eta_o ( \partial_x^2 v_x -\partial_x \partial_y v_y  +  \partial_y^2 v_x + \partial_y \partial_x v_y)
% \end{bmatrix}
% \end{align}

%%%%%%%%%%%%%%%%%%%%%%%%%%%%%%%%%%%%%%%%%%%%%%%%%%

% The system in non-dimensionalised 
% to reduce the number of free variables. In choosing the capillary timescale given
% \begin{align}
%     t_{\gamma} =  \sqrt{\frac{\rho L^3}{\gamma}}
% \end{align}
% the timescale over which capillary events occur has been normalized. This removes the sensitivity of absolute values from the equation of motion and instead highlights the relevance of the ratio between capillary and viscous or inertial effects independent of system size.
% To achieve the dimensionless equations, the variables are substituted with 
The equations are implemented in non-dimensional form. We use the capillary timescale to nondimensionalize time and the wavelength of the initial wave $L=\lambda$ to nondimensionalize length. With these choices, our variables are rescaled according to
\begin{align}
    % & \rho \frac{Du}{Dt} =  - \nabla p + \eta_e \nabla^2 u + \eta_o \epsilon \nabla^2 u +f_{\sigma}+f_{g}\\
    &\vec{x} = L\,\bar{\vec{x}},  \quad t =  \sqrt{\frac{\rho L^3}{\gamma}}\bar{t}, \quad  \vec{u}= \sqrt{\frac{\gamma}{\rho L}}\bar{\vec{u}}, \quad p =  \frac{\gamma}{L}\bar{p}, \quad 
    \vec{\sigma}^{\even, \odd} = \mu^{\even, \odd}\sqrt{\frac{\gamma}{\rho L^3}}\bar{\vec{\sigma}}^{\even, \odd}, \quad
    \kappa =  \frac{1}{L}\bar{\kappa}, \quad \delta_s =  \frac{1}{L}\bar{\delta}_s.  \label{def:DS params}
\end{align}
% The notation here has changed as a variable $A$ can be represented by  $a A'$ where $a$ in this example contains the unit of scale and $A'$ the dimensionless variable.

Substituting these re-scalings into the momentum balance leads to the following nondimensional equation of motion
% \begin{align}
%      &\frac{Du'}{Dt'} =  - \nabla p' + \frac{\eta_e}{\sqrt{\rho \gamma L}} \nabla^2 u' + \frac{\eta_o}{\sqrt{\rho \gamma L}} \epsilon \nabla^2u'
%      % + \kappa \delta_s\boldsymbol{\hat{n}} 
% \end{align}
% Which can be represented by two dimensionless quantities as
\begin{align}
    \bar{\rho} \frac{D\bar{\vec{u}}}{D\bar{t}} =  - \nabla \bar{p} + \nabla \cdot \left( \bar{\mu}^\even \, \Oe \, \bar{\vec{\sigma}}^\even \right)  + \nabla \cdot \left( \bar{\mu}^\odd \, \Oo \, \bar{\vec{\sigma}}^\odd \right) + \bar{\kappa} \, \bar{\delta}_s \, \vec{n},
    % + \kappa \delta_s\boldsymbol{\hat{n}}
    % & Bo= \frac{\rho g L^2}{\sigma} \\
    % \Oe = \frac{\eta_e}{\sqrt{\rho \sigma L}}
\end{align}
where the nondimensional density and viscosity VOF functions are now given by
\begin{align}
\bar{\rho} (c) & =  c + (1 - c)\rho_a/\rho_w, \\
\bar{\mu}^\even (c) & = c + (1 - c)\mu_a^\even/\mu_w^\even, \\
\bar{\mu}^\odd(c) & = c,
\end{align}
the Ohnesorge numbers are defined as
\begin{align}
    \Oe = \frac{\mu_e}{\sqrt{\rho \gamma L}}, \quad \Oo = \frac{\eta_o}{\sqrt{\rho \gamma L}},
\end{align} 
and surface tension is implemented as an interfacial body force ($\bar{\kappa} \, \bar{\delta}_s \, \vec{n}$) where $\bar{\delta}_s$ is the Dirac delta function centered on the interface.

% The Ohnesorge number (can be seen as dimensionless viscosity) represents the strength of viscous forces to surface tension and inertia. 
% (also given by $Oh = \frac{\sqrt{We}}{Re}$). 
% Ohnesorge numbers between $0.001 <Oh<0.01$ typically have a viscosity similar to rain while Ohnesorge numbers between $0.01<Oh<0.1$ encompass fluids from ocean sprays and agricultural sprays  \cite{sanjay2025unifying}. 

For the simulations with a sine-wave initialization, we define the nondimensional domain as a square $[0, 4] \, \times \, [0, 4]$ with periodic boundary conditions on the left and right sides. The fluid interface is initialized with the shape
\begin{equation}
h_0(x) = H_0 + a\sin{ \left( \frac{2\pi}{\lambda}x \right) },
\label{eq:initial_shape_sine}
\end{equation}
where we keep fixed $\lambda = 1$ to guarantee our results are nondimensionalized with the wavelength, and $H_0 = 3$ which satisfies the deep-water assumption. The amplitude $a$ is varied through the paper.

The fluid is initialized at rest and the mesh is adaptively refined over time as the interface deforms under surface tension and viscous forces. This refinement is always maximized near the interface and coarsened away from it, as illustrated in Fig.~\ref{fig:mesh} with Basilisk cell levels going from 5 to 9. These levels correspond to cell sizes  going from $\Delta_{max} =0.125$ to $\Delta_{min} = 0.0078125$.

\section{Dispersion relations of Odd capillary waves}
\label{app:Dispersion}
For the bulk of the fluid the linearized equations read 
\begin{gather}
    \diver \vec{u}=0\\
    \ptderiv{\vec{u}}=-\grad P+\Oe\nabla^ 2\vec{u}%-g\hat{\vec{y}}
\end{gather}
with $\vec{v}\doteq (u,v) $ and the corrected pressure $P=p-\Oo\omega $ from te vorticity $\omega=|\rot \vec{u}|$. Using the corrected pressure allows us to just use the Lamb solutions for the velocity and pressure fields

\begin{equation}
\begin{aligned}
u & =\left(A|k| e^{|k| y}+B m e^{m y}\right) e^{i k x-i \varpi t} \\
v & =-i k\left(A e^{|k| y}+B e^{m y}\right) e^{i k x-i \varpi t} \\
P & =\varpi \frac{k}{|k|} A e^{|k| y} e^{i k x-i \varpi t}%-g y
\end{aligned}
\end{equation}
with $m^2=k^2-i\varpi/\Oe$ which yields a vorticity 
\begin{equation}
\omega =e^{i k x-i \varpi t}\left(\frac{i \varpi}{\nu} B e^{m y}\right) \\
\end{equation}
thus the hydrostatic pressure is
\begin{equation}
    p=P+\Oo\omega =\varpi \left( \frac{k}{|k|} A e^{|k| y}+\frac{i \Oo}{\nu} B e^{m y} \right ) e^{i k x-i \varpi t}%-g y
\end{equation}
These solutions need to obey both the 
kinematic boundary condition (KBC)
\begin{equation}
\vec{v}=\vec{u}|_{\partial D}  ,  
\end{equation}
and the dynamic boundary condition (DBC)
\begin{equation}
\vec{\sigma}\hat{\vec{n}}|_{\partial D}=\kappa\hat{\vec{n}}.    
\end{equation}
Taking $h$ as the height of the perturbation of the surface for small amplitude waves, we can approximate
\begin{equation}
    \begin{gathered}
        \hat{\vec{n}}\approx \hat{\vec{y}}-\pderiv{h}{x}\hat{\vec{x}}\\ 
        \kappa\hat{\vec{n}}\approx \frac{\partial^2 h}{\partial x^2}\hat{\vec{y}}
     \end{gathered}
\end{equation}

Moreover, from the KBC one gets 
\begin{equation}
    \ptderiv{h}+u\pderiv{h}{x}=v|_{\partial D\equiv y=0}
\end{equation}
that in linearizing and in Fourier space yields 
\begin{equation}
   %-i\varpi h =-i k\left(A e^{|k| y}+B e^{m y}\right) e^{i k x-i \varpi t} \Big|_{y=0} \Rightarrow 
   h = \frac{k}{\varpi}\left(A+B \right) e^{i k x-i \varpi t}
\end{equation}
for the DBC we need to write down the stress tensor $\vec{\sigma}$. 
\begin{equation}
    \sigma_{ij}=-p\delta_{ij}+\Oe \left(\partial_iu_j+\partial_ju_i\right)-{\Oe\partial_ku_k\delta_{ij}}+\Oo \left(\partial_i\varepsilon_{jk}u_k+\varepsilon_{ik}\partial_ku_j\right)
\end{equation}
where we can cancel out the compression term. \begin{comment}Explicitly we have 
\begin{equation}
\vec{\sigma}=-p\begin{pmatrix}1&0\\0&1
\end{pmatrix}+\nu \begin{pmatrix}
2 \partial_x u & \partial_y u + \partial_x v \\
\partial_y u + \partial_x v & 2 \partial_y v
\end{pmatrix} +\nu_o\begin{pmatrix}
\partial_y u + \partial_x v &  \partial_y v -\partial_x u \\
 \partial_y v -\partial_x u  & -\partial_y u - \partial_x v
\end{pmatrix}
\end{equation}
that we can transform to 
\begin{equation}
\vec{\sigma}=-P\begin{pmatrix}1&0\\0&1
\end{pmatrix}+\nu \begin{pmatrix}
2 \partial_x u & \partial_y u + \partial_x v \\
\partial_y u + \partial_x v & 2 \partial_y v
\end{pmatrix} +\nu_o\begin{pmatrix}
2\partial_y u  &  -2\partial_x u \\
 2\partial_y v   &  - 2\partial_x v
\end{pmatrix}
\end{equation}
again with $P=p-\nu_o\varpi$ and using the incompressibility condition to simplify the off-diagonal terms of the last matrix 
\end{comment}
As such, the DBC leads to 
\begin{align}
    \Oe(\partial_y u+\partial_x v)+2\Oo\partial_x u&=0\\
    -P+2\Oe\partial_y v+2\Oo\partial_x v&= \partial_{xx}h
\end{align}

which, substituting the Lamb solutions, discarding gravity, yields
\begin{multline}
    \begin{pmatrix}
\varpi^2- |k|^3+2i k(\Oe k  -i  \Oo |k|)\varpi  &2ik( m\operatorname{sgn }(k)\nu -i|k| \Oo)\varpi  -|k|^3 \\
 -\varpi^2+|k|^3    &\varpi^2 + |k|^3 -2ik \big[ (  m\operatorname{sgn}(k)-k)\Oe   +   i(m-|k|)\Oo \big]\varpi
    \end{pmatrix}
\begin{pmatrix}iA-B|k|/k\\
iA+B|k|/k
    \end{pmatrix}=0.
\end{multline}
Which, leads to the characteristic equation, in non-dimensional form of 
\begin{equation}
    \begin{vmatrix}
 \varpi^2+2 \, ik \, \big( \Oe  \, k-i  \, \Oo|k|\big) \, \varpi -|k|^3  & 2 \, ik \, \big( \Oe  \,  m \, \operatorname{sgn}k-i  \, \Oo|k|\big) \, \varpi  -|k|^3\\[1em]
  2 \, ik \, \big( \Oe \,  k-i  \, \Oo |k|\big)  & \varpi   + 2 \, ik \, \big( \Oe k-i \,  \Oo m\big)
    \end{vmatrix}=0
\end{equation}

%\begin{multline}
%    \left(|k|-\sqrt{k^2-\frac{i \varpi }{\nu }}\right) \left(4 i k^3 \varpi  \left(\nu ^2+\nu_o^2\right)-2 i \gamma  \nu_o k^4\right)+\\+i \gamma  k^3
%   \varpi -2 i \nu_o |k| \varpi^2 \sqrt{k^2-\frac{i \varpi }{\nu }}+\frac{|k| \varpi ^2 \left(4 \nu % k^2-i \varpi \right)}{k}-2 i \nu_o
%   k^2 \varpi ^2=0
%\end{multline}
%
In the limit of zero viscosities, $\Oe\to0$ and $\Oo\to0$, we recover the known dispersion 
\begin{equation}
    \varpi^2=|k|^3
\end{equation}
If instead we take the limit of zero shear viscosity $\Oe\to0$ one gets

\begin{equation}
\varpi^2 +2 \Oo k|k| \varpi -  |k|^3  =0
\end{equation}

having the asymptotic behaviour 
\begin{equation}
    \varpi=\frac{1}{2\Oo}k \quad\text{and}\quad \varpi=-2\Oo k|k|
\end{equation}
as $\Oo\sim1$. The next order correction on the even viscosity brings up the dissipative nature of the modes as 
\begin{equation}
    \varpi=\frac{k}{2\Oo}\bigg(1-2i\frac{\Oe}{{\Oo}}\bigg) \quad\text{and}\quad \varpi=-2\Oo k|k|\bigg(1-\frac{i}{2}\sqrt{\frac{\Oe}{\Oo}}\operatorname{sgn}k\bigg)
\end{equation}

\subsection*{Influence of gravity on the dispersion}
While deriving the dispersion relation and throughout the discussion, we discarded the effects of gravity on the system. We can justify this approach on both thematic and technical grounds. Regarding the latter, the computation of the dispersion with gravity effects is almost trivial, as one can replace $|k|^3\mapsto|k|^3+{\rm Bo}|k|$ and resort to the determinant of the sum of $2\times2$ matrices to obtain 
\begin{equation}
\varpi^2 +2 \Oo k|k| \varpi -  |k|^3  -{\rm Bo}|k|=0
\end{equation} 
in the same low viscosity limit as before and where ${\rm Bo}=\rho g L^2/\gamma$ is the Bond number. Thus, the inclusion of gravity effects simply symmetrically deforms the branches and does not play a role in the nonreciprocal phenomena that we are interested in studying. Beyond that, it amounts to a sublinear correction that only becomes noticeable for substantial values of Bond number and quite long wavelengths. 

\begin{figure}[ht!]
    \centering
    \includegraphics[width=0.5\linewidth]{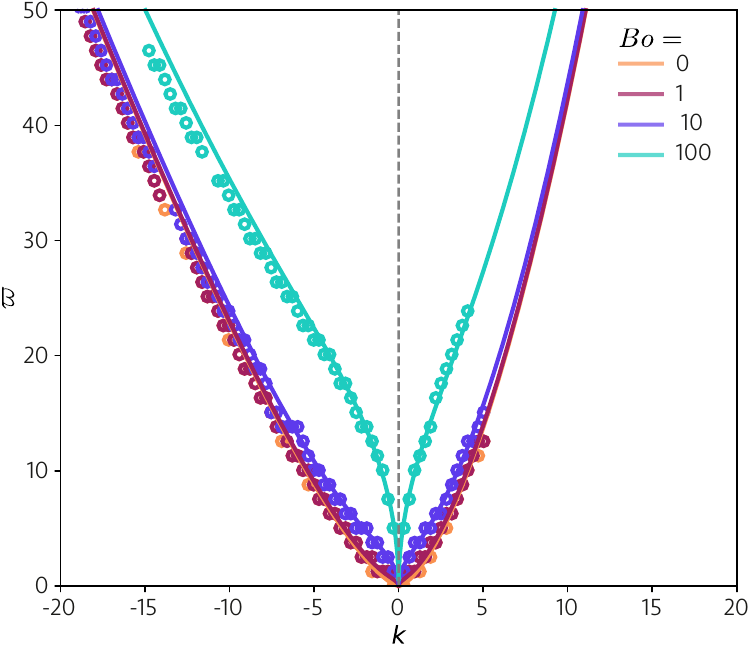}
    \caption{Dispersion relation, $\varpi(k)$, for different Bond numbers, $Bo$; solid lines show the linear theory and circles the numerical solutions.}
    \label{fig:gravitycapwaves}
\end{figure}

Yet, for the sake of completeness, we tested a number of simulations that include gravity, and we still obtain good agreement with our linear theory as can be observed in Fig.~\ref{fig:gravitycapwaves}, where it is also shown that the deformation of the branches calls for ${\rm Bo}\approx100$.

\section{Minimal nonlinear model of capillary waves}
\label{app:NonlinearModel}
Our simulations show that nonlinear effects are critical to understand the full behaviour of the surface perturbations. As such, we aim to get a minimal weakly nonlinear scalar model that could go further on the theoretical treatment of the system. 

We require such a model to recover the same dispersion relation in its linear limit. So, we start by inverting the dispersion $\varpi^2 +2 \Oo k|k| \varpi -  |k|^3  =0$ from reciprocal to real space. There, the treatment of the absolute value of $k$ requires us to resort to the Hilbert transfrom, defined as the multiplier operator with symbol $-i\,{\rm sign}(k)$ such that $\widehat {\operatorname{H}[\phi]}=-i\,{\rm sign}(k)\,\widehat{\phi}$ that leads to $\widehat {\operatorname{H}[\partial_x\phi]}=|k|\,\widehat{\phi}$ and that can be written as the convolution operator with a Cauchy kernel,  
\begin{equation}
\operatorname{H}[\phi](t) = \frac{1}{\pi}\,\fint_{-\infty}^{+\infty} \frac{\phi(\tau)}{t - \tau}\,\mathrm{d}\tau.
\end{equation}
Thus, the inversion of the dispersion relation leads us to the linear nonlocal PDE of the form
\begin{equation}
    \phi_{tt}-\operatorname{H}\left[2\Oo \phi_{xxt}+\phi_{xxx}\right]=0,
\end{equation}
for a scalar field $\phi$ representing the amplitude of the eigenvectors of the linearly perturbed system. To introduce the nonlinearity, we aim for a convective-like term $\sim\phi\phi_x$ (that can also be interpreted as a first term in the expansion of a more complex nonlinear coupling) and moreover, such that in the limit of $\Oo\gg1$ one recovers a classical Benjamin--Ono equation~\cite{benjamin1967internal,ono1975algebraic}, previously also proposed for odd surface waves without surface tension \cite{Abanov_2018}. With the mentioned constraints, our model reads 
\begin{equation}
    \phi_{tt}+(s \phi\phi_x)_t-\operatorname{H}\left[2\Oo \phi_{xxt}+ \phi_{xxx}\right]=0\label{eq:nonlinear2ndBOApp}
\end{equation}
where we introduce the coupling constant $s$. Note that, in a general case $s$ can be a function of $\Oo$ and, more critically, of the propagation velocity of the wave and be different for right and left propagating modes, further breaking reciprocity (a similar behaviour was studied in the context of self-interacting odd fluids in \cite{Cosme2023a}) but in order to keep the model at its maximal simplicity we will treat it as a uniform constant.

Looking for traveling solutions of the form $\phi(\xi)\equiv \phi(x-ct)$ transforms  \eqref{eq:nonlinear2ndBOApp} into 
\begin{equation}
    c^2\phi''-cs(\phi\phi')'-(1-2\Oo c)\operatorname{H}\left[\phi'''\right]=0
\end{equation}
and one can find periodic solutions of the form 
\begin{equation}
    \phi(\xi)=\frac{2c}{s}\frac{\tanh^2\alpha}{1{\pm}\operatorname{sech}\alpha \cos\frac{\pi\xi}{L}}
\end{equation}
where the parameter $\alpha$ is defined through
\begin{equation}
\tanh\alpha=   \frac{\pi(1-2\Oo c)}{c^2L}.
\end{equation}
In the limit of infinite wavelength $L\to\infty$ we then recover the algebraic soliton
\begin{equation}   
\phi(\xi)=\frac{4c(1-2\Oo c)^2{s^{-1}}}{c^4\xi^2+(1-2\Oo c)^2}.
\end{equation}

%%%%%%%%%%%%%%%%%%%%%%%%%%%%%%%%%%%%%%%%%%%%%%%%%%

% \section{Additional Material}
% \begin{figure}
%     \centering
%     \includegraphics[width=00.8\linewidth]{figures/appendix_ksin.png}
%     \caption{Caption}
%     \label{fig:enter-label}
% \end{figure}
% \begin{figure}
%     \centering
%     \includegraphics[width=00.8\linewidth]{figures/appendix_phase.png}
%     \caption{Caption}
%     \label{fig:enter-label}
% \end{figure}

% \begin{figure}
%     \centering
%     \includegraphics[width=0.6\linewidth]{figures/stokes_drift_A0.05.png}
%     \caption{Caption}
%     \label{fig:enter-label}
% \end{figure}
% \begin{figure}
%     \centering
%     \includegraphics[width=0.8\linewidth]{figures/appendix_kbox.png}
%     \caption{Caption}
%     \label{fig:enter-label}
% \end{figure}

\begin{comment}

\begin{figure}[p]
    \centering
    \includegraphics[width=0.9\linewidth]{figures/appendix/Odd_kymo_APP.pdf}
    \caption{Kymographs for A0.05,A0.2}
    \label{fig:placeholder1}
\end{figure}
\begin{figure}
    \centering
    \includegraphics[width=0.9\linewidth]{figures/appendix/odd_ddm_app.pdf}
    \caption{Dispersion Kymographs A0.1}
    \label{fig:placeholder2}
\end{figure}

\begin{figure}[p]
    \centering
    \includegraphics[width=0.5\linewidth]{figures/appendix/odd_delta_app.pdf}
    \caption{Dispersion Kymographs A0.1}
    \label{fig:placeholder3}
\end{figure}
\end{comment}

\end{document}

%% file: initialisation.tex
\usepackage{titlesec}
\titleformat{\section}
  {\bf\sffamily}
  {\thesection. }
  {5pt}
  {\MakeUppercase}
\renewcommand{\thesection}{\Roman{section}} 

\titleformat{\subsection}
  {\bf\sffamily}
  {\thesubsection. }
  {5pt}{}
  
\renewcommand{\thesubsection}{\Alph{subsection}}

\usepackage[affil-it]{authblk} 
\usepackage{etoolbox}
\usepackage{lmodern}

\usepackage{eurosym}

\usepackage[utf8]{inputenc}
\usepackage{geometry}
\usepackage[hidelinks]{hyperref}
\usepackage[citestyle=numeric,style=phys,biblabel=brackets,backend=bibtex,sorting=none,url=false,doi=false, natbib]{biblatex}
\bibliography{bibliography}
\DefineBibliographyStrings{english}{andothers={\itshape et\addabbrvspace al\adddot}}

\usepackage{csquotes}
\usepackage[]{authblk} 
\usepackage{etoolbox}
\usepackage{lmodern}

\usepackage{amsmath}
\usepackage{enumerate}
\usepackage{enumitem}
\usepackage{graphicx}
\usepackage{siunitx}
\usepackage{float}
\usepackage[]{parskip} %skip=0.7em, indent=1.5em

\makeatletter
\patchcmd{\@maketitle}{\LARGE \@title}{\fontsize{16}{19.2}\selectfont\@title}{}{}
\makeatother

\usepackage[normalem]{ulem}

\usepackage{graphicx}% Include figure files
\usepackage{dcolumn}% Align table columns on decimal point
\usepackage{bm}% bold math
\usepackage{tikz}
\usetikzlibrary{math}
\usetikzlibrary{shapes.geometric, arrows}
\usetikzlibrary{positioning}
\usetikzlibrary{shapes,arrows}
\usetikzlibrary{intersections}
\usepackage{csvsimple}
\usepackage{changepage}
\usepackage[tbtags]{mathtools}
\usepackage{siunitx}
\usepackage{csquotes}

\usepackage{float}
\usepackage{subfigure}
\usepackage[
	nonumberlist, 				% keine Seitenzahlen anzeigen
	acronym,      				% ein Abk�rzungsverzeichnis erstellen
	nomain,						% no main glossary
	nopostdot,					% Kein punkt nach Beschreibung
]{glossaries}
\usepackage{glossary-superragged}
\usepackage[hidelinks]{hyperref}
\usepackage{color, colortbl}

\usepackage{wrapfig}

\usepackage{pgfplots}
\usepackage{tikz}
\usetikzlibrary{arrows}
\usepackage{amsmath}
\pgfplotsset{compat=newest}
\usepgfplotslibrary{fillbetween}
\usetikzlibrary{calc}
\def\centerarc[#1](#2)(#3:#4:#5)% Syntax: [draw options] (center) (initial angle:final angle:radius)
{ \draw[#1] ($(#2)+({#5*cos(#3)},{#5*sin(#3)})$) arc (#3:#4:#5); }

\usepackage{amssymb}
\let\vec\mathbf
\usepackage{nicefrac}

\usepackage{abstract}
\usepackage{multirow}
\usepackage{tabularx}
\usepackage{booktabs}
\usepackage{array}
\usepackage{longtable}
\newcolumntype{L}[1]{>{\raggedright\let\newline\\\arraybackslash\hspace{0pt}}m{#1}}
\newcolumntype{C}[1]{>{\centering\let\newline\\\arraybackslash\hspace{0pt}}m{#1}}
\newcolumntype{R}[1]{>{\raggedleft\let\newline\\\arraybackslash\hspace{0pt}}m{#1}}

\newacronym{3d}{3D}{three dimensional}
\newacronym{am}{AM}{additive manufacturing}
\newacronym{fdm}{FDM}{fused deposition modeling}
\newacronym{ism}{ISM}{in-space manufacturing}
\newacronym{iss}{ISS}{International Space Station}
\newacronym{fcb}{FCB}{Functional Cargo Block}
\newacronym{dem}{DEM}{discrete element method}
\newacronym{md}{MD}{molecular dynamics}
\newacronym{dc}{DC}{direct-current}
\newacronym[plural=PFCs,firstplural=parabolic flight campaigns (PFCs)]{pfc}{PFC}{Parabolic Flight Campaign}
\newacronym{fft}{FFT}{Fast Fourrier Transform}
\newacronym{cad}{CAD}{Computer Assisted Design}
\newacronym{ptfe}{PTFE}{polytetrafluoroethylene}
\newacronym{ps}{PS}{polystyrene}
\newacronym{nasa}{NASA}{National Aeronautics and Space Administration}
\newacronym{esamm}{ESAMM}{Extended Structure Additive Manufacturing Machine}
\newacronym{amf}{AMF}{Additive Manufacturing Facility}
\newacronym{us}{US}{United States}
\newacronym{usa}{USA}{United States of America}
\newacronym{bmgs}{BMGs}{Bulk Metallic Glasses}
\newacronym{esa}{ESA}{European Space Agency}
\newacronym{si}{SI}{International System of Units, abbreviated from French \textit{Syst\`{e}me International (d'unit\'{e}s)}}
\newacronym{dlr}{DLR}{German Aerospace Center}
\newacronym{liggghts}{LIGGGHTS}{\acrshort{lammps} Improved for General Granular and Granular Heat Transfer Simulations}
\newacronym{lammps}{LAMMPS}{Large-scale Atomic/Molecular Massively Parallel Simulator}
\newacronym{sjkr}{SJKR}{Simplified Johnson-Kendall-Roberts}
\newacronym{ded}{DED}{Directed Energy Deposition}
\newacronym{slm}{SLM}{Selective Laser Melting}
\newacronym{sls}{SLS}{Selective Laser Sintering}
\newacronym{eva}{EVA}{Extra-Vehicular Activity}
\newacronym{sem}{SEM}{Scanning Electron Microscopy}
\newacronym{RPM}{RPM}{Ramdom Positioning Machine}
\newacronym{rpm}{rpm}{revolutions per minute}
\newacronym{rise}{RISE}{Research Internships in Science and Engineering}
\newacronym{daad}{DAAD}{German Academic Exchange Service, abbreviated from German \textit{Deutscher Akademischer Austauschdienst}}
\newacronym{fsm}{FSM}{finite-state machine}
\newacronym{ir}{IR}{infrared}
\newacronym{pcbs}{PCBs}{Printed Circuit Boards}
\newacronym{pcb}{PCB}{Printed Circuit Board}
\newacronym{mcr}{MCR}{Modular Compact Rheometer}
\newacronym{sff}{SFF}{Solid Freeform Fabrication}
\newacronym{uv}{UV}{ultraviolet}
\newacronym{abs}{ABS}{acrylonitrile butadiene styrene}
\newacronym{hpde}{HPDE}{high density polyethylene}
\newacronym{pei}{PEI}{polyetherimide}
\newacronym{bff}{BFF}{BioFabrication Facility}
\newacronym{lens}{LENS}{Laser Engineered Net Shaping}
\newacronym{cnc}{CNC}{Computer Numerical Control}
\newacronym{ebf3}{EBF$^3$}{Electron Beam Free-Form Fabrication}
\newacronym{leo}{LEO}{Low Earth Orbit}
\newacronym{pc}{PC}{polycarbonate}
\newacronym{crissp}{CRISSP}{Customisable Recyclable International Space Station Packaging}
\newacronym{Athena}{Athena}{Advanced Telescope for High-ENergy Astrophysics}
\newacronym{lbm}{LBM}{Laser Beam Melting}
\newacronym{bam}{BAM}{Federal Institute for Materials Research and Testing, abbreviated from German \textit{Bundesanstalt f\"{u}r Materialforschung und-pr\"{u}fung}}
\newacronym{pbf}{PBF}{powder bed fusion}
\newacronym{eb}{EB}{Electron Beam}
\newacronym{2d}{2D}{two dimensional}
\newacronym{4d}{4D}{four dimensional}
\newacronym{ft4}{FT4}{Freeman Technology 4 Powder Rheometer}
\newacronym{dsc}{DSC}{Differential Scanning Calorimetry}
\newacronym{pmma}{PMMA}{polymethylmethacrylate}
\newacronym{1g}{$1g$}{gravity on-ground}
\newacronym{mug}{$\mu g$}{microgravity}
\newacronym{bcm}{BCM}{Box Counting Method}
\newacronym{mct}{MCT}{Mode Coupling Theory}
\newacronym{gmct}{gMCT}{granular Mode Coupling Theory}
\newacronym{itt}{ITT}{Integration Through Transients}
\newacronym{mfc}{MFC}{Mass Flow Controller}
\newacronym{ct}{CT}{computed tomography}
\newacronym{xct}{XCT}{X-ray computed tomography}
\newacronym{cv}{CV}{curriculum vitae}
\newacronym{pi}{PI}{principal investigator}
\newacronym{osp}{OSP}{orthogonal superimposed perturbation}
\newacronym{npi}{NPI}{Network Partnering Initiative}
\newacronym{ecsat}{ECSAT}{European Centre for Space Applications and Telecommunications}
\newacronym{eac}{EAC}{European Astronaut Centre}
\newacronym{estec}{ESTEC}{European Space Research and Technology Centre}
\newacronym{fps}{fps}{frames per second}
\newacronym{pdf}{pdf}{probability density function}
\newacronym{al}{Al}{aluminium}
\newacronym{ss}{\textit{SS}}{\textit{Smooth Surface}}
\newacronym{rs}{\textit{RS}}{\textit{Rough Surface}}
\newacronym{rcp}{rcp}{random close packing}
\newacronym{iop}{IoP UvA}{Institute of Physics of the University of Amsterdam}
\newacronym{mp}{MP}{Institute of Material Physics for Space}
\newacronym{elgra}{ELGRA}{European Low Gravity Research Association}
\newacronym{zarm}{ZARM}{Center of Applied Space Technology and Microgravity}
\newacronym{piv}{PIV}{particle image velocimetry}

%% file: colors_define.tex
\definecolor{brickred}{rgb}{0.8, 0.25, 0.33}
\definecolor{darkorange}{rgb}{1.0, 0.55, 0.0}
\definecolor{persiangreen}{rgb}{0.0, 0.65, 0.58}
\definecolor{persianindigo}{rgb}{0.2, 0.07, 0.48}
\definecolor{cadet}{rgb}{0.33, 0.41, 0.47}
\definecolor{turquoisegreen}{rgb}{0.63, 0.84, 0.71}
\definecolor{sandybrown}{rgb}{0.96, 0.64, 0.38}
\definecolor{blueblue}{rgb}{0.0, 0.2, 0.6}
\definecolor{ballblue}{rgb}{0.13, 0.67, 0.8}
\definecolor{greengreen}{rgb}{0.0, 0.5, 0.0}